\documentclass[epj]{svjour}

\usepackage{graphics}
\usepackage{epsfig}
\usepackage{amssymb}
\begin{document}
\title{Multi-scale correlations in different futures markets}

\author{M. Bartolozzi\inst{1,2}, C. Mellen\inst{1}, T. Di Matteo\inst{3}, T. Aste\inst{3}}
\offprints{marco.bartolozzi@gmf.com.au}          
\institute{Research Group, Grinham Managed Funds, Sydney NSW 2065,
Australia \and Special Research Centre for the Subatomic Structure
of Matter (CSSM), University of Adelaide, Adelaide SA 5005,
Australia \and Department of Applied Mathematics, Research School
of Physical Sciences and Engineering,
 The Australian National University, Canberra ACT 0200, Australia}

\abstract{In the present work we investigate the multiscale nature of the
correlations for high frequency data (1 minute) in different
futures markets over a period of two years, starting on the
$1^{\rm{st}}$ of January 2003 and ending on the $31^{\rm{st}}$ of
December 2004. In particular, by using the concept of {\em local}
Hurst exponent, we point out how the behaviour of this parameter,
usually considered as a benchmark for persistency/antipersistency
recognition in time series, is largely time-scale dependent in the
market context. These findings are a direct consequence of the
intrinsic complexity of a system where trading strategies are
scale-adaptive. Moreover, our analysis points out different
regimes in the dynamical behaviour of the market indices under
consideration.}
\PACS{
      {}{Econophysics}   \and
      {}{Multiscale Phenomena}   \and
      {}{Detrended Fluctuation Analysis} \and
      {}{Time series analysis}
} 
\maketitle
\section{Introduction: persistency and anti-persistency in non-stationary systems}
\label{sec::introduction}

In recent years we have witnessed the development of a new branch
of research on the edge between physics and economics. This new
area, nowadays widely recognized in both the communities, goes
under the name of {\em econophysics}. One of the most important
achievement of this novel discipline has been to point out
empirically that the stock market is far from being efficient:
memory processes and feedbacks are present and they play a quite
important role in the dynamics of this system.
In particular, several studies have addressed the analysis of
market fluctuations or {\em logarithmic returns}, defined as
$r(t)=\ln [ P_{r}(t)/P_{r}(t-1) ]$, where $P_{r}(t)$ is the price
of a certain market at time $t$. Interestingly, the results show
that the shape of the probability distribution function (pdf),
$P$, irrespective of the particular stock under consideration,
displays a leptokurtic behaviour\footnote{The actual shape of the
distribution of returns is still a matter of debate. Intriguing
frameworks have been recently proposed by Tsallis~\cite{Tsallis03}
and Beck~\cite{Beck02}. A more complete discussion on this
important topic is beyond the scope of the present work.}, that is
``fat" tails, whose asymptotic decay can be well approximated by a
power law, $P(r) \sim r^{-\beta}$,  with exponent $ \beta \sim 3$.
This result is very important and in fact openly contrasts with
the standard assumption that for a long time has ruled the
academic world of theoretical economics, that is, the {\em
efficient market hypothesis} (EMH)~\cite{Dacorogna01b}.
 According to the EMH the
dynamics of market price movements are equivalent to that of white
noise and, therefore, their pdf can be well represented by a
Gaussian. In other words, the very large fluctuations observed in
the empirical price movement distribution, and represented by the
power law tails, should not exist (statistically).
For a broader discussion on this subject and the field of
econophysics the interested reader can refer to the books and
reviews in
Refs.~\cite{Mantegna99,Bouchaud99,Paul99,Dacorogna01a,Feigenbaum03,Proceedings06,Proceedings07}.

The source of the ``anomalous" behaviour in the market dynamics
has to be related to {\em inefficiencies}, such as  feedbacks in
the price which, eventually, lead to very large fluctuations, such
as crashes. It is obvious that the exploitation of these
inefficiencies, even if for limited periods of time, becomes
extremely important for traders and financial companies.

For a single asset, inefficiencies are also related to
correlations in the price value over time. It is well known that
first order or linear correlations can be neglected for most of
the indices when looking at time scales longer than a few
minutes~\cite{Mantegna99,Bouchaud99}. This does not rule out the
possibility of higher order correlations, but, in order to extract
these, we need to make use of tools that are more sophisticated
than the standard autocorrelation function. Moreover, we need to
consider possible non-stationarities that may affect the time
series: the dynamics of the stock market behaves differently
according to different ``environmental'' conditions such as, for
example, changes in the market regulation or in the trading
mechanism itself.

{\em Detrended fluctuation analysis} (DFA), recently proposed by
Peng at al.~\cite{Peng94} in the context of  DNA nucleotides
sequences, has been developed in order to extract correlations
from time series with local trends - that is, from non-stationary
times series. This method is particulary relevant not only to
finance but also to areas such as geophysics or biophysics - where
non-stationarity is the rule rather than the exception. The DFA
method, summarized in Sec.~\ref{sec::DFA}, is based on the
calculation of the average variance related to a certain trend at
different scales. This procedure leads to an estimation, via a
scaling relation, of the {\em Hurst exponent}, $H \in [0,1]$, of
the time series: for $0 \le H < 0.5$ it is said that the behaviour
of the time series is {\em antipersistent}, and conversely, {\em
persistent} for $0.5 < H \le 1$. For completely uncorrelated
movements, as assumed by the EMH, we expect $H=0.5$.  Note that
the idea of calculating persistency/antipersistency in time series
through the scaling of the variance is not peculiar to the DFA but
in fact dates back to the pioneering work of Hurst (and so we
obtain the name Hurst exponent) in the context of reservoir
control on the Nile river dam project, around
1907~\cite{Feder88,Hurst51}.

In the present work we investigate the temporal evolution at
different scales of the {\em local} Hurst
exponent~\cite{Muniandy01,Costa03,Cajueiro04,Grech04,Carbone04},
where by the term ``local'' we indicate the Hurst exponent
calculated at time $t$ over a certain temporal window $L$ that
extends backward in time. This concept is very important for
non-stationary and multiscale systems such as the stock market.
Here, the dynamics of the trading can be influenced at different
horizons by differences in the portfolio of strategies used by
traders. In this case there is no reason to believe that $H$
should remain the same for all $t$ or that it would not vary if we
calculated it using windows of a different length. For this reason
the Hurst exponent is considered as ``local'', in both time and
length scale.

In Sec.~\ref{sec::DFA_pdfs} we back up the previous arguments with
an empirical analysis where we show, using pdfs of local Hurst
exponents, that correlations depend not only on the particular
period under consideration but also on the length scale that we
are observing, therefore confirming the multiscale nature of the
market dynamics. Moreover, we point out how this technique can be
used to monitor changes in the dynamics of the market which can be
clearly observed for some specific indices.

Despite the simplicity of the previous arguments, extracting a
value of the local Hurst exponent that accurately describes the
serial correlation in a time series is not a trivial task.
Important limitations arise from a number of different sources. A
first limitation comes from the fixed temporal scale, $L$, that
sets the limit for the number of data points to be used in the
analysis. Usually a reliable estimation of $H$ requires a large
number of samples which, on the other hand, prevents the
investigation of very small scales. A second limitation is related
to the possible presence of non-Gaussian increments in the time
series. Large non-stationary increments are able to make
significant contributions to the observed value of $H$. In this
situation the range spanned by the variance over a time interval
may not be related to a sequence of temporally correlated steps in
a certain direction but instead may be primarily determined by
large and possibly self-similar jumps that are temporally
uncorrelated.

Furthermore, we have to consider the intrinsic precision of the
algorithm used to calculate $H$: that has to be considered as
another source of uncertainty. These issues are discussed in
Sec.~\ref{sec::preliminary}.
In this regard it is important to stress that, in recent years,
there has been a proliferation of methods devoted to the
estimation of $H$ or of scaling exponents in general. To the best
of our knowledge each method is characterized by a unique set of
advantages and disadvantages. The choice of the DFA as the working
tool in the present work is related mainly to its vast popularity
and, therefore, to the necessity of properly understanding its
limitations. For a recent review on scaling methods in finance the interested reader
could refer to \cite{DiMatteo07}.

The financial time series used in the analysis presented in Secs.~\ref{sec::preliminary} and
\ref{sec::DFA_pdfs}, are composed of 1 minute prices for different futures contracts starting from
January 2003 up until the end of December 2004. In particular we analyze:

\begin{itemize}
    \item Stock Indices: Dax (DA), Euro Stoxx (XX), Standard \& Poor500 (SP), Dow (DJ), Hang Seng
    (HI) and Nikkei255  (NK).
    \item Commodities: Gold COMEX (GC) and Crude Oil E-mini (QM).
    \item Exchange Rates: Japanese Yen (JY) and British Pound
    (BP).
    \item Fixed Income: Eurex Bunds (BN), Long Gilts (GL), Treasury Bonds (US) and BOBL (BL).
\end{itemize}

Each data set contains approximately $3 \cdot 10^5$ samples,
depending on the specific contract.

In summary, the present work is organized as follows: in Sec.~\ref{sec::DFA}
we describe the algorithm used for the calculation of the
local Hurst exponent, that is the DFA. In
Sec.~\ref{sec::preliminary}, we show some possible pitfalls of
this method for ``fat" tailed time series by using {\em fractional
Brownian motion} and {\em L$\acute{\rm e}$vy processes} as working
examples. The main analysis is presented in
Sec.~\ref{sec::DFA_pdfs} while discussions and conclusions are
left for the last section.

\section{The detrended fluctuation analysis method}
\label{sec::DFA}

DFA, originally proposed in Ref.~\cite{Peng94}, is considered one
of the most powerful technique to extract correlations
from non-stationary time series. This peculiarity makes the DFA
suitable for applications to stock market time series. Some
examples of its use in this field are given in
Refs.~\cite{Cizeau97,Liu97,Vandewalle97,Liu99,Janosi99,Gopikrishnan00,Gopikrishnan01,Muniandy01,Matia02,Costa03,Grech04,Ivanov04,Eisler06}.

 The main idea behind this method is to analyze
 the scaling of the average fluctuations around a possible
deterministic local trend of some sort. In practice, if we have a
time series of random movements in time (in the same fashion as a
random walk), $x(t)$, of total length $N$, which in the stock
market case can be identified with the logarithmic price, $x(t)
\equiv \ln[P_{r}(t)]$, then the implementation of the method can
be summarized as following:

\begin{enumerate}

\item {The time series is divided in $M=N/\tau$ non-overlapping
boxes of equal length $\tau$. In this case $x^{i}_{\tau}(t)$
represent the sub-series of length $\tau$ associated with the
$i^{th}$ box.}

\item{For each box, first we calculate the local trend which, it
is assumed, can be approximate by a polynomial of degree $p$,
$y^{i}_{\tau,p}(t)$, and then the fluctuations, $F^{i}(\tau,p)$,
around their local trend as

\begin{equation}
F^{i}(\tau,p)=\sqrt{\frac{1}{\tau} \sum_{t \in i^{th} \,
box}(x^{i}_{\tau}(t)-y^{i}_{\tau,p}(t))^{2}}.
\end{equation}

According to the order of the polynomial used for the detrending,
we indicate DFA as DFA-$p$. }

\item{ As the final step, the average fluctuation over the $M$
boxes $\langle F(\tau,p) \rangle_{M} = (1/M)
\sum_{i=1}^{M}F^{i}(\tau,p)$ is calculated along with the
$\tau$-scaling which, for a certain range of values, behaves as a
power law

\begin{equation}
\langle F(\tau,p)\rangle_{M} \propto \tau^{H}.
 \label{eq::pow_law}
\end{equation}

From  Eq.(\ref{eq::pow_law}) we can finally extract the scaling
exponent, for example via a linear fit over the scales where the
power law holds. }

\end{enumerate}

 Note that the accuracy of
the method can be slightly influenced by the order of the
polynomial used for the detreding and in principle it would be
more correct to write $H \equiv H(p)$. However for the clarity of
notation we will not write the parameter $p$ as a variable for
$H$. We will return to this matter in the next section where a
 study on the dependency of $H$ on $p$ has been carried
out.

As mentioned in the introduction, the dynamics of the stock market
can shift between phases of persistency and anti-persistency and
hence there is no reason, {\em a priori}, to believe that the
Hurst exponent has to remain constant over long periods of time.
In this context it is useful to introduce the concept of {\em
local} Hurst exponent, that is, the Hurst exponent calculated at a
certain time $t \le N$ over a time window $L \ll N$ which extends
backwards from $t$. The choice of the window length $L$ is very
important from a theoretical point of view. In fact, the value of
$H$ in Eq.~(\ref{eq::pow_law}), as we will see in the next
section, can change according to choice of $L$ - hence $L$ can be
identified as a characteristic time for our calculations. The
different behaviour of the Hurst exponent at different scales is
nothing but an expression of the multiscale dynamics of the system
enhanced by the averaged coarse-grained procedure used by the DFA
algorithm when calculating the scaling relation
Eq.~(\ref{eq::pow_law}). Thus the Hurst exponent at a certain
temporal scale $L_{0}$ can be different from the one calculated at
a different scale $L_{1}$, where, for example, $L_{1} \gg
L_{0}$\footnote{Note that if the EMH is realized then $H = 0.5$
should hold independently on the particular period of time, $t$,
or on the particular window, $L$, apart from numerical
inaccuracies, of course.}.
 In order to stress these dependencies
of the parameter $H$ we denote the local Hurst exponent calculated
at a certain scale $L$ by $H \equiv H_{L}(t)$.
If we now successively shift the time window $L$ by a discrete time lag
$\Delta t$ we are able to construct a time series of local Hurst
exponents and so monitor the dynamics of the system during its
evolution by extracting useful information on the correlation at a
particular scale and in a particular period of time, as shown in
Figs. \ref{fig::ts-DFA-1}, \ref{fig::ts-DFA-2} and
\ref{fig::ts-DFA-3} of Sec.~\ref{sec::DFA_pdfs}.

However we are not completely free in our choice of $L$, being this parameter
bounded by computational limitations mainly related to the minimum
 number of points needed to have a reliable calculation of the
 value of $H$. So far, not many studies have been devoted
 to this issue \cite{Weron02,Xu05} and usually, in practical applications,
 a rule of thumb is used. To fill this gap, in the next section we carry out an
 analysis devoted to investigate possible drawbacks and pitfalls of
 the DFA-$p$ method with variable $L$, in particular when applied to ``fat" tailed data sets,
 as the case of the data investigated in the present work.

\section{DFA: application to data sets with ``fat" tails}
\label{sec::preliminary}

In order to be able to perform a reliable analysis on our sets of
financial data we must have an idea of the accuracy of the DFA-$p$
method under the conditions that we are going to use it.

Since we are interested in studying the changes in the correlation
at short scales it would be appropriate to have an estimation of
the error for small data sets. In fact, it is well known that,
when dealing with power law relations such as
Eq.~(\ref{eq::pow_law}), the most accurate results are achieved
when a very large sample of data, spanning over several scales, is
available.

Moreover, as we mentioned in the introduction, high frequency
financial time series are characterized by large, non-Gaussian,
fluctuations that are responsible for the ``fat" tailed shape of
the pdf of the returns~\cite{Dacorogna01a}. This feature can
provide an important contribution to the value of $H$ even if
there is no serial correlation (persistency/antipersistency) at
all among the data. Some implications of broad tailed data in the
multifractal contest and for the R/S algorithm have been discussed
in Refs.~\cite{Kantelhardt02} and \cite{Alfi07} respectively.

\subsection{DFA with Gaussian increments}
\label{sec::gaussian_inc}

 Before tackling the important issue of
the ``fat" tails and analyzing financial data, let us here give an
estimation of the error associated with the calculation of $H$ in
case of Gaussian increments. In particular, we test the DFA-$p$
algorithm against an ensemble of 500 short time series ($L =
1024$) of {\em fractional Brownian motion} (FBM) with tunable $H$,
generated via a wavelet-construction method (WFBM) described in
Ref.~\cite{Abry96}.
 The average values of $H$, along with the standard deviations, are reported in
 Tab.~\ref{tab::averages} for DFA-1 and DFA-2.

\begin{table}
\caption{Values of Hurst exponent evaluated for the WFBM with DFA-1
and DFA-2 corresponding to the nominal value reported in the first
column. In both cases, $p=1$, and quadratic, $p=2$,
polynomials to estimate the local trend. In call cases $L =
1024$.}\label{tab::averages}.
\begin{tabular}{|c||c|c|}
  \hline
    & DFA-1 & DFA-2 \\
  \hline
  \hline
  $H=0.2$ & $0.22 \pm 0.03$ & $0.22 \pm 0.04$ \\
  $H=0.3$ & $0.30 \pm 0.04$ & $0.31 \pm 0.04$ \\
  $H=0.4$ & $0.40 \pm 0.05$ & $0.40 \pm 0.04$ \\
  $H=0.5$ & $0.50 \pm 0.06$ & $0.50 \pm 0.05$ \\
  $H=0.6$ & $0.60 \pm 0.07$ & $0.60 \pm 0.06$ \\
  $H=0.7$ & $0.70 \pm 0.08$ & $0.70 \pm 0.06$ \\
  $H=0.8$ & $0.79 \pm 0.08$ & $0.79 \pm 0.07$ \\
  \hline
\end{tabular}

\end{table}

\begin{figure}
\vspace{1cm}
\centerline{\epsfig{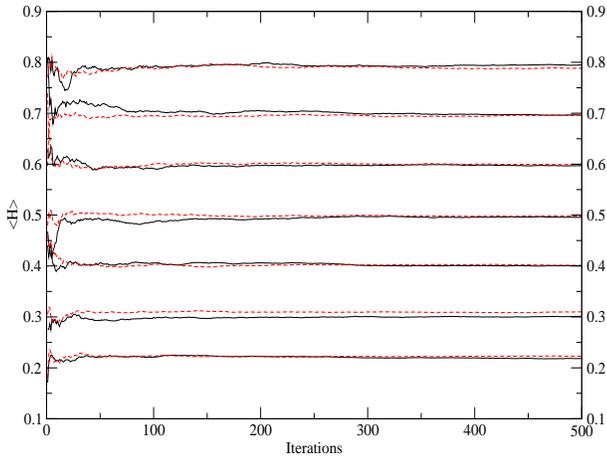}}
\caption{Time series of the average value of $H$, $\langle H
\rangle$, as a function of the number of members in the WFBM
ensemble, each with length $N=1024$. The fact that each time
series is relatively short prevents in-sample self-averaging from
occurring and it actually takes the addition of a reasonable
number of ensemble members before $\langle H \rangle$ converges to
a stable value. The continuous lines refer to DFA-1 while the
dashed ones to DFA-2.} \label{fig::DFA12}
\end{figure}

The two examples give very similar results. Note that we obtain a
slightly biased estimate of $H$ for high values of correlation,
($H \gtrsim 0.8$), and anticorrelation, ($H \lesssim 0.3$). The
systematic errors are toward smaller and larger $H$, respectively.
However, these values are highly unrealistic for correlations in
market movements where we expect to find $H$ not too far from 0.5
(this would be different in case of volatilities or
volumes~\cite{Liu99,Ivanov04,Eisler06b}). Note also that standard
deviations of the DFA-2 $H$ estimates are generally smaller than that
returned by DFA-1 and, therefore, we will use DFA-2 to carry out
our analysis from now on. However, we must stress that the results
reported in Sec.~\ref{sec::DFA_pdfs} are not influenced by the
particular choice of $p$. In Fig.~\ref{fig::DFA12} we show the
average values of $H$ as a function of the number of members in
the WFBM ensemble.

\begin{figure}
\vspace{1cm}
\centerline{\epsfig{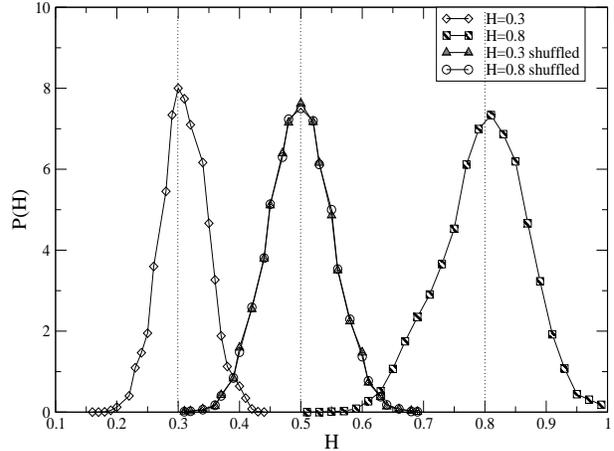}}
\caption{Pdfs for $H$ for the correlated data set of WFBM with
Gaussian increments and $L=1024$. As expected, the distributions
for the original data sets are centered in H=0.3 and H=0.8. Note
also that the distributions are slightly skewed toward 0.5 leading
to a slightly biased estimation of $H$ for these high correlations
- as shown also in Fig.~\ref{fig::DFA12}. The distribution of the
$H$ values for the shuffled data is instead a Gaussian centered on
0.5. These pdfs have been averaged over 5 independent shuffling.}
\label{fig::fbm}
\end{figure}

In Fig.~\ref{fig::fbm} we report for DFA-2 the pdfs obtained by
the previous 500 ensembles for the specific cases of $H=0.3$ and
$H=0.8$. From this plot we can clearly see the source of the bias,
pointed out in the previous paragraph, as being a skewness toward
0.5 in the distributions of the evaluated Hurst exponents. Note
also that if we randomly shuffle the time series, as shown in the
same plot, the resulting pdfs are perfectly symmetric and centered
around 0.5, as expected for uncorrelated increments. For $0.3
\lesssim H \lesssim 0.7$, the distributions of Hurst exponents,
not shown, are Gaussian distributed and centered on the
``expected" nominal value. Therefore, we can deduct that these
increments do not lead any systematic bias in the DFA-2 algorithm,
at least for values of $H$ included in the range previously
mentioned. In all cases the statistical uncertainty can be
considered as a good estimate of the error in the $H$ estimate.
The source of the small systematic error revealed for $H \gtrsim
0.8$ and $H\lesssim 0.3$ is not clear - it could be in the
generating process or in the estimation algorithm. It will not be
considered further at this time since we are extremely unlikely to
face these kinds of correlations when studying price movements.
However, the source of this bias may be of interest for studies
where high (low) values of the Hurst exponent are involved - such
as for volatilities and volumes.

\begin{figure}
\vspace{1cm}
\centerline{\epsfig{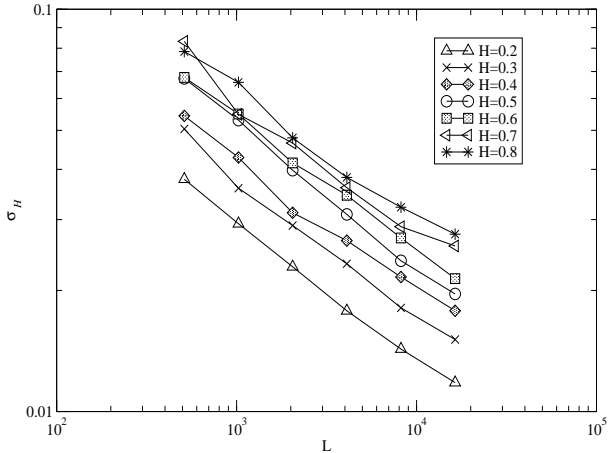}}
\caption{The standard deviation of estimated $H$ against $L$ for
Gaussian increments from the WBFM process. We can notice a power
law relation with exponent $\sim 0.36$, independent of $H$. Note
also that for a fixed $L$ the standard deviation grows with $H$. }
\label{fig::sigmaH_L}
\end{figure}

Once again using ensembles of time series produced using the WBFM
method, we can try to understand how the statistical error in our
estimate of $H$ varies with $L$. Results from this study are
plotted in Fig.~\ref{fig::sigmaH_L}. From this figure we deduce
that the standard deviation $\sigma_{H} \propto L^{-\gamma}$, with
$\gamma \sim 0.36$, irrespective of the particular value of $H$
considered. This scaling law is a consequence of self-averaging
effects for stationary time series. Apart from this, and as
previously noticed, the absolute value of the error slightly grows
with $H$.


\subsection{DFA with L$\acute{\rm e}$vy increments}
\label{section:DFA with Levy}

Now we turn our attention to the challenges posed by the large
fluctuations which characterize high frequency financial time
series. As a first step we consider the relationship between the
Hurst exponent and i.i.d. increments $y$, generated at a temporal
scale $\tau$ by a symmetric {\em $\alpha$-stable L$\acute{e}$vy
process}~\cite{Samorodnitzky94,Paul99}. For $\alpha \in (0,2)$ the
pdf of these increments is characterized by fat tails and a
probability distribution
\begin{equation}
L_{\alpha}(y,\tau) \sim \frac{1}{|y|^{1+\alpha}},
\end{equation}
for $y \rightarrow \pm \infty$. As a consequence, the variance and
all higher moments are infinite. Moreover, the pdf of a
L$\acute{\rm e}$vy process obeys the scaling property
\begin{equation}
L_{\alpha}(y,\tau)=\tau^{-\frac{1}{\alpha}}
L_{\alpha}(\tau^{-\frac{1}{\alpha}}y,1) \equiv
\tau^{-\frac{1}{\alpha}}
L^{\star}_{\alpha}(\tau^{-\frac{1}{\alpha}}y),
\label{eq::levy_dist}
\end{equation}
which is equivalent to saying that, irrespective of the
observation scale of the process, we can always rescale the
increments and the time so that every observed pdf can be
collapsed into a pdf, $L^{\star}_{\alpha}$, of rescaled increments
$\tau^{-1/\alpha}y$ and $\tau=1$. This feature characterizes the
statistical self-similarity of the process, despite the fact that
the variables are independent.

The same feature is found for self-affine increments $x(t)$
defined as
\begin{equation}
x(\lambda t) = \mu(\lambda) x(t), \label{eq::self-similar}
\end{equation}
to which the Hurst exponent is related\footnote{Note that in Eq.
(\ref{eq::self-similar}) the equal sign holds in the statistical
sense.}~\cite{Samorodnitzky94,Groenendijk98,Kiyani06}. In fact, if
the pdf of $x(t)$ at certain scale $t$ is known - say $P(x(t),t)$
- then we can derive the corresponding pdf for the rescaled
variable in $\tau = \lambda t$, $G(x(\lambda t),\lambda t)$. From
simple probability considerations we have that
\begin{equation}
G(y, \tau)= \left| \frac{dx}{dy}\right| \, P(x,t)\big
\vert_{y,\tau},
\end{equation}
where $y=x(\lambda t)$. It follows that
\begin{equation}
G(y, \tau)=\frac{1}{\mu(\lambda)} \, P \left (\frac{y}
{\mu(\lambda)}, \frac{\tau}{\lambda} \right ),
\end{equation}
and, since the rescaling factor is arbitrary, setting $\lambda
=\tau$ gives that
\begin{equation}
G(y, \tau)=\frac{1}{\mu(\tau)} \, P \left (\frac{y}{\mu(\tau)}, 1
\right ) \equiv \frac{1}{\mu(\tau)} \, P^{\star}
\left(\frac{y}{\mu(\tau)} \right ), \label{eq::self-affine}
\end{equation}
where $P^{\star}$ is the collapsing pdf with time increments
$\tau=1$. For mono-fractal self-affine time series the Hurst
exponent is defined as $\mu(\lambda)=\lambda^{H} \equiv \tau^{H}$.
By comparing Eq.~(\ref{eq::levy_dist}) and
Eq.~(\ref{eq::self-affine}) we can deduce that $H=1/\alpha$. For
$0<\alpha<2$ we are in a fat tail regime where $H>0.5$ is related
just to the self-affinity of the increments and not to serial
correlations. Note that  $\alpha=2$ is a special case of stable
distribution, i.e., the Gaussian. In this case the Hurst exponent
assumes the well known value for uncorrelated signals, $H=0.5$.
For an exhaustive discussion of L$\acute{\rm e}$vy and
self-similar processes we refer the reader to the book of
Samorodnitzy and Taqqu~\cite{Samorodnitzky94}.

In order to validate the results presented in the previous
discussion and to test the reliability of DFA-2 on L$\acute{\rm
e}$vy processes we have performed numerical tests for different
values of $H>0.5$ (since $\alpha \in (0,2)$) and different lengths
$L$ of the time series. The L$\acute{\rm e}$vy increments $y$ have
been generated via the algorithm proposed in reference
\cite{Samorodnitzky94}. Accordingly

\begin{equation}
y=\frac{\sin (\alpha \phi)}{[\cos(\phi)]^{1/\alpha}} \, \left [
\frac{\cos ((1-\alpha)\phi)}{\nu}
\right]^{\frac{1-\alpha}{\alpha}},
\end{equation}

where $\phi$ is a uniformly distributed random number in the
interval $[-\pi/2,\pi/2]$ and $\nu$ an independent realization of
an exponential random variable with mean 1. The results for the
average value of $H$ are reported in Tab.~\ref{tab::levy} for the
average obtained using 500 realizations of the L$\acute{\rm e}$vy
process of length $L$\footnote{Note that, since the second moment
of the distribution diverges, this procedure is just limited to
finite time series.}. In this case, with the process increments
being serially independent a shuffling of them would have no
effect and hence all the contributions to $H$ come from the
self-similarity of the increments. Note that the accuracy of the
DFA-2 algorithm in this case is quite poor. This result is not
surprising at all: DFA-$p$ has been developed specifically for
particular kind of trends - those that can be well described via a
polynomial - and not for the large (spiky) jumps such as the ones
considered in this example~\cite{Hu01}.

\begin{table}
\caption{Values of Hurst exponent evaluated for L$\acute{\rm e}$vy
processes of different $H=1/\alpha$ and and $L$.}\label{tab::levy}
\begin{tabular}{|c||c|c|c|}
  \hline
    & $L=1024$ & L=4096 & L=16384\\
  \hline
  \hline
  $H=0.6$ & $0.56 \pm 0.08$ & $0.58 \pm 0.07$ & $0.57 \pm 0.05$\\
  $H=0.7$ & $0.63 \pm 0.10$ & $0.64 \pm 0.08$ & $0.65 \pm 0.07$\\
  $H=0.8$ & $0.68 \pm 0.12$ & $0.70 \pm 0.10$ & $0.71 \pm 0.08$\\
  \hline
\end{tabular}

\end{table}

Although the tails for the L$\acute{\rm e}$vy stable processes
tend to be systematically broader that the ones of the financial
data~\cite{Mantegna99,Bouchaud99} we can use the results from
these preliminary studies as a general indication of the likely
behaviour of DFA-$2$ when applied to ``fat" tailed data.
Accordingly it seems not unreasonable to expect that the presence
of large non-Gaussian fluctuations in stock market data will give
rise to a bias toward large $H$. This contribution is of no help
in inferring serial correlation and, therefore, should be taken
into consideration when drawing conclusions on the
persistency/antipersistency of a time series.

\subsection{DFA for futures indices}

\begin{figure}
\vspace{1cm}
\centerline{\epsfig{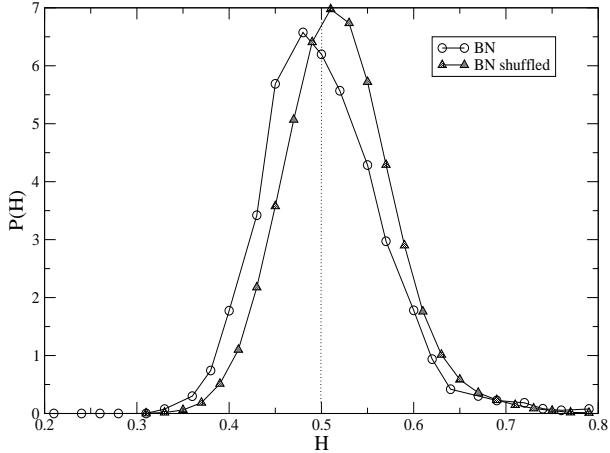}}
\caption{Pdfs of $H_{L}(t)$ for the BN time series with $L=1024$.
Note that the distribution of the original data is peaked on a
value slightly below 0.5. On the other side, the shuffled curve
is not peaked around 0.5 but at a higher value.}
\label{fig::BN_pdfs}
\end{figure}

In this section we apply the DFA-2 to different sets of financial data.
Let us start considering the logarithmic price of the 1 minute BN (Bund futures)
time series. At this time scale large fluctuations are very
frequent and the pdf of returns displays a pronounced leptokurtic
shape. In this case, instead of producing an ensemble of short
time series, we fix a time frame $L$ and we slide it through the
all the data set at intervals of 10 minutes. In addition, we applied
the DFA-2 estimator to sets of shuffled BN data. The pdfs for the
raw and shuffled data sets are shown in Fig.~\ref{fig::BN_pdfs}.

The results are quite different to those obtained from i.i.d
Gaussian increments. At this scale, $L=1024$, the 1 minute BN time
series shows on average a slightly antipersistent behaviour. The
shuffled set, on the other hand, displays a slightly persistent
behaviour. This latter observation can be seen as an indication of
the systematic contribution toward large $H$ values that may be
expected when large non-Gaussian fluctuations are present in the
time series. Such an outcome was in fact presaged by the
discussion in section \ref{section:DFA with Levy}. Results from
the other sets of futures data show equivalent behaviour. Hence,
it can perhaps be argued that if one is interested in determining
$H$ attributable to serial correlations in a data set then one
should also take into account the ``offset'' in $H$ that can arise
from any large non-Gaussian fluctuations that may be present. This
offset may be estimated by evaluating the
persistency/antipersistency of the shuffled data.

\begin{figure}
\vspace{1cm}
\centerline{\epsfig{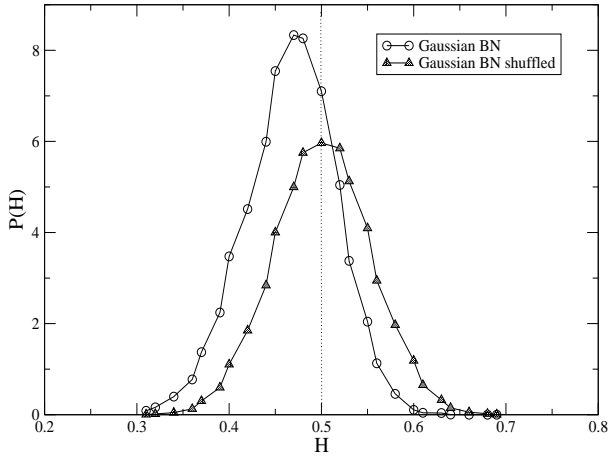}}
\caption{Pdfs of $H_{L}(t)$ for the surrogate BN time series with
$L=1024$. The distribution of the surrogate data is peaked on a
value slightly smaller than the one found for the original set,
underlying the positive contribution of the large fluctuations.
Moreover, the shuffled version is this time centered on 0.5.}
\label{fig::gs_BN_pdfs}
\end{figure}

In order to double check that the source of the persistence in the
shuffled BN time series is related to the ``fat" tails of the data
set we have created a surrogate time series of the BN returns
($r$) data according to

\begin{equation}
\left \{  \begin{array}{ccc}
r(t) \rightarrow |g(t)|& {\rm if} & r(t)>0,\\
r(t) \rightarrow -|g(t)|& {\rm if} & r(t) < 0,
\end{array} \right.
\end{equation}

where $g(t)$ is a random Gaussian increment. In this way, while
keeping the possible temporal correlations in the increments
direction, we get rid of the large fluctuations that, as we saw in
the previous subsection, can be a possible source of an unwanted
contribution. The results of the analysis for the BN surrogates
are shown in Fig.~\ref{fig::gs_BN_pdfs}. In this case, the pdf
displays anticorrelation, in a similar fashion to the original
time series but with a peak that is centered at a slightly lower
value of $H$. More interesting, the Hurst exponent for the
shuffled version of the surrogates is centered around 0.5 as it
should be for non-correlated time series. These are important
results - in fact we have shown that the value of $H$ obtained
from short leptokurtic sets is, in reality, the result of two
contributions: the possible genuine temporal correlations and the
self-similarity in the non-Gaussian increments, as observed for
L$\acute{\rm e}$vy processes.

For the 1 min time series that we use in the present work we give
an estimate of the average contribution to $H$ due to the large
self-similarity fluctuations: this can change from time series to
time series according to their degree of intermittency. In
particular we report in Tab.~\ref{tab::bias}, for each index and
different $L$, the mean value of $H$ after shuffling the
increments along with the relative difference from the theoretical
value 0.5. The former can change from values of approximately 15\%
for the most ``bursty" indices as NK, JY or BP to 2\% for SP and
DJ. Note also the small sensitivity on the window size for some
indices. This effect is related to the phenomenon of {\em
volatility clustering}~\cite{Mantegna99,Bouchaud99} which enhances
the value of $H$ when smaller windows are considered.

\begin{table*}
\caption{The average values of $H$ along with the standard
deviations are reported for different indices and scales after an
average over 5 independent shuffling of the time series. The
relative difference, $\Delta H /H$, from the expected value of 0.5
is also given. } \label{tab::bias}
\begin{tabular}{|c||c|c|c|c|c|c|c|}
  \hline
  \hline
  & $L=512$ & $L=1024$ & $L=2048$ & $L=4096$ & $L=8192$ & $L=16384$ & $\Delta H /H$ \\
  \hline
  SP & $0.51\pm 0.08$ & $0.51\pm 0.05$ & $0.51\pm 0.04$ & $0.51\pm 0.03$ & $0.51\pm 0.03$ & $0.51\pm 0.02$ &  $\approx$ 2\%  \\
  DJ & $0.51\pm 0.08$ & $0.51\pm 0.05$ & $0.51\pm 0.04$ & $0.51\pm 0.03$ & $0.51\pm 0.03$ & $0.51\pm 0.02$ &  $\approx$ 2\%  \\
  NK & $0.58\pm 0.10$ & $0.58\pm 0.07$ & $0.57\pm 0.05$ & $0.56\pm 0.04$ & $0.56\pm 0.03$ & $0.55\pm 0.02$ &  $\approx 16\div 10$\% \\
  HI & $0.55\pm 0.09$ & $0.55\pm 0.06$ & $0.55\pm 0.05$ & $0.54\pm 0.04$ & $0.54\pm 0.03$ & $0.53\pm 0.02$ &  $\approx 10\div 6$\% \\
  DA & $0.52\pm 0.09$ & $0.52\pm 0.06$ & $0.52\pm 0.04$ & $0.51\pm 0.03$ & $0.51\pm 0.03$ & $0.51\pm 0.02$ &  $\approx 4\div 2$ \%  \\
  XX & $0.51\pm 0.08$ & $0.51\pm 0.05$ & $0.51\pm 0.04$ & $0.51\pm 0.03$ & $0.51\pm 0.03$ & $0.51\pm 0.02$ &  $\approx$ 2\%  \\
  \hline
  GC & $0.53\pm 0.09$ & $0.53\pm 0.05$ & $0.52\pm 0.04$ & $0.52\pm 0.03$ & $0.52\pm 0.03$ & $0.52\pm 0.02$ &  $\approx 6\div 4$\%  \\
  QM & $0.53\pm 0.09$ & $0.53\pm 0.06$ & $0.52\pm 0.04$ & $0.52\pm 0.03$ & $0.52\pm 0.03$ & $0.52\pm 0.02$ &  $\approx 6\div 4$\%  \\
  \hline
  JY & $0.56\pm 0.10$ & $0.56\pm 0.06$ & $0.56\pm 0.05$ & $0.55\pm 0.04$ & $0.55\pm 0.03$ & $0.54\pm 0.02$ &  $\approx 12\div 8$ \% \\
  BP & $0.57\pm 0.10$ & $0.57\pm 0.07$ & $0.56\pm 0.05$ & $0.56\pm 0.04$ & $0.55\pm 0.03$ & $0.54\pm 0.02$ &  $\approx 14\div 8$ \% \\
  \hline
  BN & $0.52\pm 0.09$ & $0.52\pm 0.06$ & $0.52\pm 0.05$ & $0.52\pm 0.04$ & $0.52\pm 0.03$ & $0.52\pm 0.02$ &  $\approx$ 4\%  \\
  GL & $0.52\pm 0.09$ & $0.53\pm 0.06$ & $0.53\pm 0.06$ & $0.53\pm 0.05$ & $0.53\pm 0.05$ & $0.53\pm 0.05$  & $\approx 6\div 4$\% \\
  US & $0.52\pm 0.09$ & $0.52\pm 0.06$ & $0.52\pm 0.05$ & $0.52\pm 0.04$ & $0.52\pm 0.03$ & $0.52\pm 0.03$ &  $\approx$ 4\% \\
  BL & $0.51\pm 0.09$ & $0.52\pm 0.06$ & $0.52\pm 0.05$ & $0.52\pm 0.04$ & $0.52\pm 0.03$ & $0.52\pm 0.03$ &  $\approx$ 4 \%  \\
  \hline
\end{tabular}

\end{table*}

It is important to underline that the estimate of the contribution
of the large fluctuations to the value of $H$, see
Tab.~\ref{tab::bias}, is just the average effect over the two years
period under consideration and cannot be directly subtracted from
the single values of the $H$ that we find. Nevertheless, it gives
an idea of the order of magnitude that this phenomenon can assume,
on average, over this time. A careful examination of the
problem would require one to consider separately each sub-window
under examination. However, the problem of manipulation of the
``outliers" is not trivial: they can contain important information
that otherwise could go missing. Moreover, we need to consider the
multiscale nature of these fluctuations: in this context a wavelet
filtering would give results appropriate for a possible
analysis~\cite{Bartolozzi06}. Part of our future work will be
devoted to a further study of this problem. The problem of
``outliers" in the calculation of a self-affine exponents with
underlying L$\acute{\rm e}$vy process has been addressed also in
Ref.~\cite{Kiyani06}.

\section{Multiscale Hurst exponent statistics from 1/1/2003 to 31/12/2004}
\label{sec::DFA_pdfs}

\begin{figure}
\vspace{1cm}
\centerline{\epsfig{figure=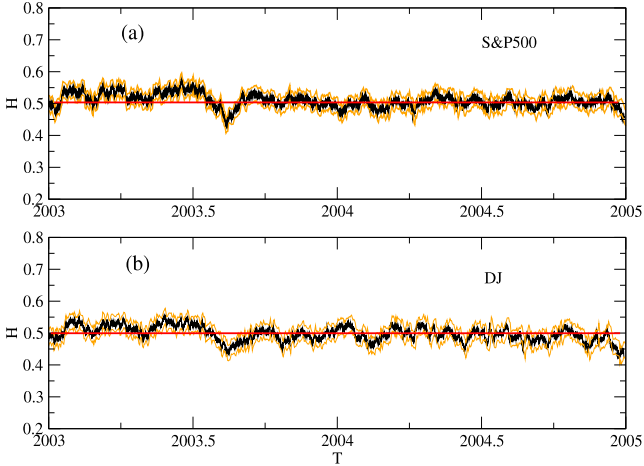,height=8cm,width=10cm}}
\caption{Time series of local Hurst exponents, $H_{L}(t)$, for the
time series SP (a) and DJ (b) on a scale of approximately 16 days
($L=8192$) and constant shift $\Delta t=10$ minutes. For these
particularly liquid markets $H$ is always very close to 0.5,
irrespective of the particular scale under consideration. A
confidence interval based on Gaussian errors is shown as well. The
time period goes from 1/1/2003 to 31/12/2004.}
\label{fig::ts-DFA-1}
\end{figure}

\begin{figure}
\vspace{1cm}
\centerline{\epsfig{figure=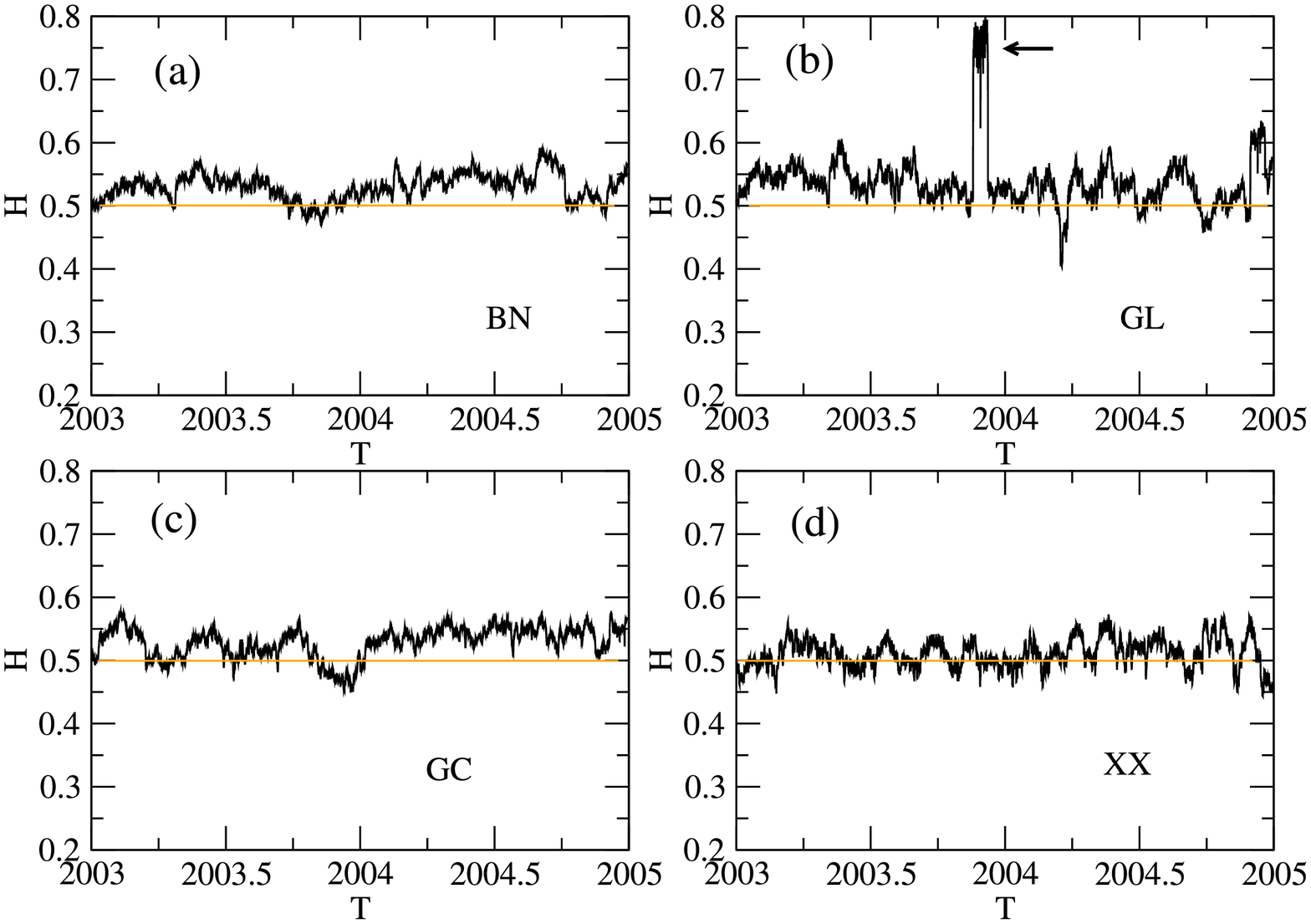,height=8cm,width=10cm}}
\caption{Local Hurst exponents with $L=8192$ and $\Delta t=10$ for
BN (a), GL (b) GC (c) and XX (d). The anomalous ``burst" observed
for GL, indicated by the arrow in (b),  is nothing but an artifact
of the data set where a large artificial gap is present, see also
Sec.~\ref{sec::preliminary}. } \label{fig::ts-DFA-2}
\end{figure}

\begin{figure}
\vspace{1cm}
\centerline{\epsfig{figure=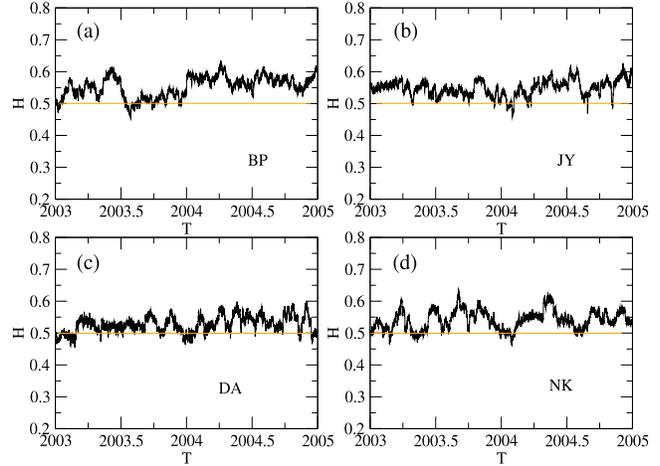,height=8cm,width=10cm}}
\caption{Local Hurst exponents with $L=8192$ and $\Delta t=10$ for
BP (a), JY (b) DA (c) and NK (d). } \label{fig::ts-DFA-3}
\end{figure}

In this section we apply the concept of local Hurst exponent, $H_L(t)$,
described in Sec.~\ref{sec::DFA} to different future indices
introduced in Sec.~\ref{sec::introduction}. The main aim here is to
investigate its global short term statistical behaviour over the
period starting from 1/1/2003 and ending 31/12/2004. The time windows
used for the analysis are the following $L=16384, 8192,4096,2048,1024$ and $512$
samples which, for the 1 minute data under consideration, span
from 32 to 1 working days approximately. For lengths shorter that
512 the computation of $H$ becomes overly contaminated by noise
and therefore we take this value as our finest scale. Examples of
time series of $H_{L}(t)$ for some indices are shown in
Figs.~\ref{fig::ts-DFA-1}, \ref{fig::ts-DFA-2} and
\ref{fig::ts-DFA-3} for $L=8192$ and a constant shift of $\Delta
t=10$ minutes.
From these figures we can notice how the Hurst exponent over
$L=8192$ minutes is not strictly stationary during the period
under consideration for any of the indices investigated. Moreover,
the dynamics of $H_{L}(t)$ appear to be quite different from index
to index. For example, the estimates of the local Hurst exponents
for the S\&P500 and the Dow Jones, shown in
Fig.~\ref{fig::ts-DFA-1}, are always very close to the value of
$0.5$ (apart from a few periods) and, considering the estimated
error over their values, there is no evidence for long periods of
persistency or antipersistency at this temporal scale. A very
similar situation is observed for the XX, Fig.~\ref{fig::ts-DFA-2}
(d). Note that SP, DJ and XX are very liquid indices with large
volumes involved: this is usually the case when markets are more
``efficient". These results are in agreement with the previous
findings in Ref.~\cite{Costa03}.


The situation looks different for other indices: well defined
periods in which $H$ significantly differs from 0.5 can be
observed. These can be due, especially for the less liquid markets
and for short periods of time, to some groups of large investors
that can alone actually drive the behaviour of the market.

\begin{figure}
\vspace{1cm} \centerline{\epsfig{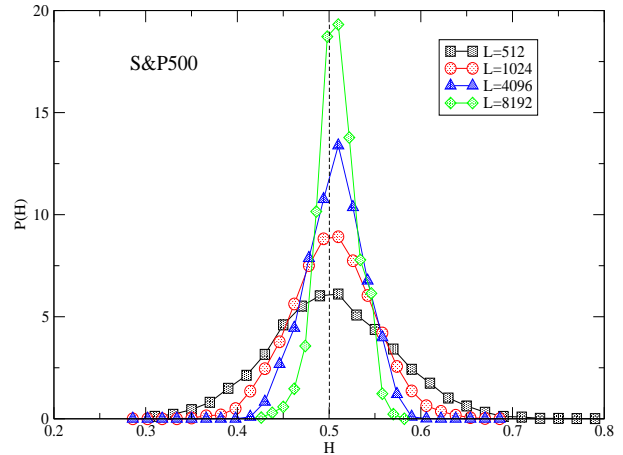}} \caption{Pdfs
for $H_{L}(t)$ at various $L$ for the S\&P500 futures. The distributions are peaked at $H \sim 0.5$
for all scales considered : the index is close to ``efficiency".} \label{fig::scaling_SP}
\end{figure}

\begin{figure}
\vspace{1cm} \centerline{\epsfig{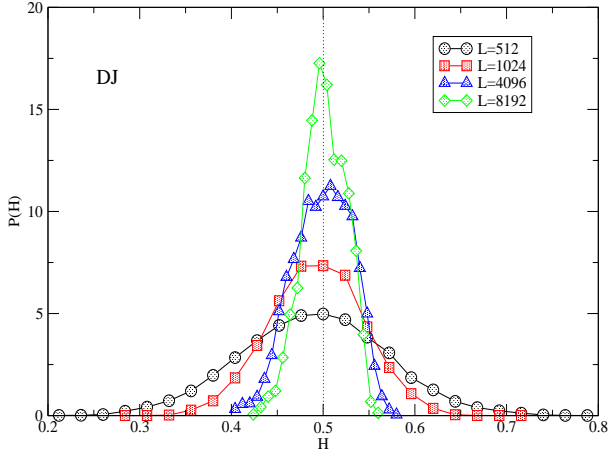}}
\caption{Similarly to those of the S\&P500 futures, the Dow Jones futures distributions show
basically no significant correlation over the time period considered. Note the shoulder for
$L=8192$: this indicates a different regime in the dynamics of the system. }
\label{fig::scaling_DJ}
\end{figure}

\begin{figure}
\vspace{1cm} \centerline{\epsfig{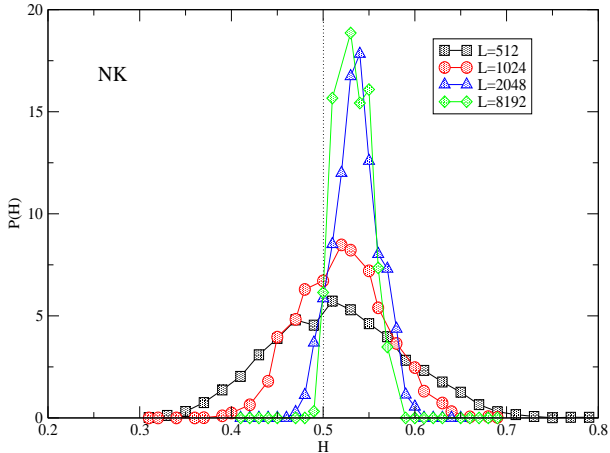}} \caption{For
the Nikkei225 futures we have generally persistent behaviour with different shoulders. }
\label{fig::scaling_NK}
\end{figure}

\begin{figure}
\vspace{1cm} \centerline{\epsfig{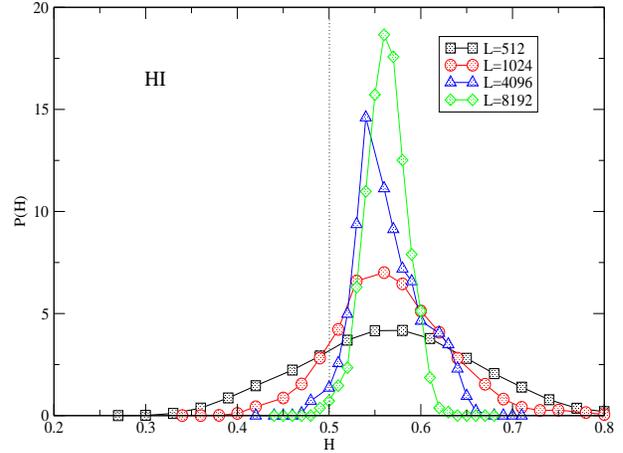}} \caption{Pdfs
for the Hang Seng futures. Interestingly this is the only index where the average dynamics drifts
toward a more persistent regime at shorter temporal scales. } \label{fig::scaling_HI}
\end{figure}

\begin{figure}
\vspace{1cm} \centerline{\epsfig{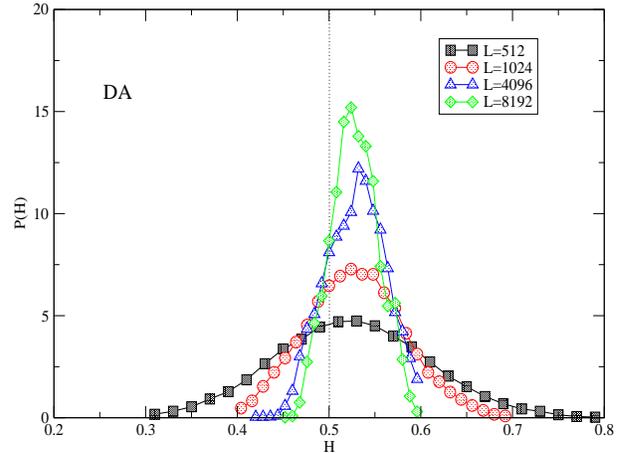}} \caption{The
German DAX futures show generally persistent behaviour with some shoulders at larger scales.}
\label{fig::scaling_DA}
\end{figure}

\begin{figure}
\vspace{1cm} \centerline{\epsfig{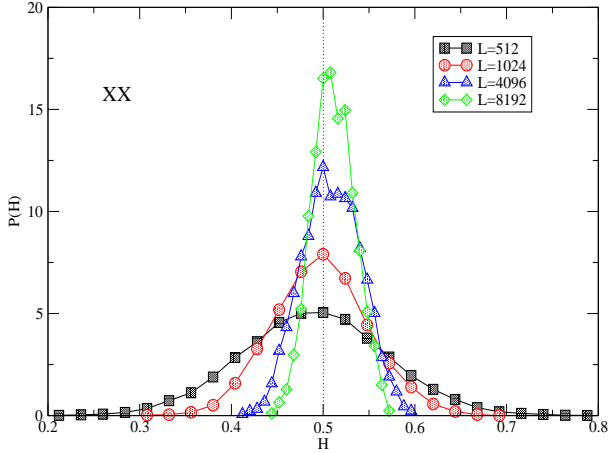}} \caption{Pdfs
for the Euro Stoxx futures. No significant average correlations can be observed for the shorter
scales. At the two longer scales shoulders are present for $H>0.5$.} \label{fig::scaling_XX}
\end{figure}

\begin{figure}
\vspace{1cm} \centerline{\epsfig{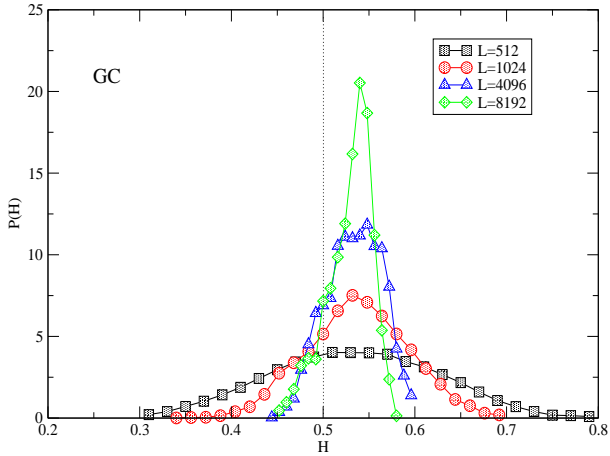}} \caption{For
Gold futures we observe persistency across all scales investigated. Noticeable shoulders are
present at a time scale of approximately eight working days ($L=4096$).} \label{fig::scaling_GC}
\end{figure}

\begin{figure}
\vspace{1cm} \centerline{\epsfig{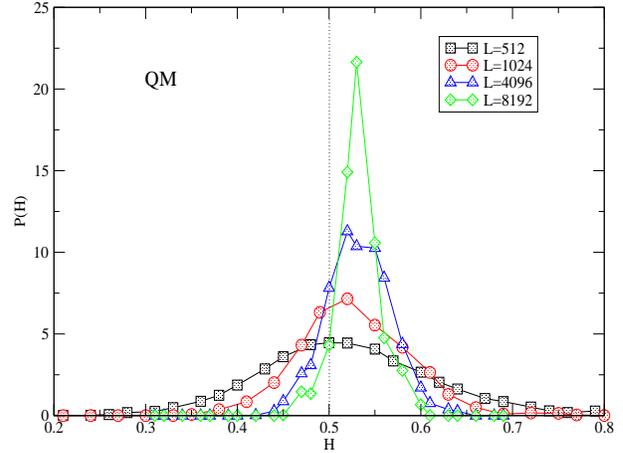}} \caption{Pdfs
for the Crude Oil futures. A shift toward a ``efficient" behaviour at short temporal scales can be
observed.} \label{fig::scaling_QM}
\end{figure}

\begin{figure}
\vspace{1cm} \centerline{\epsfig{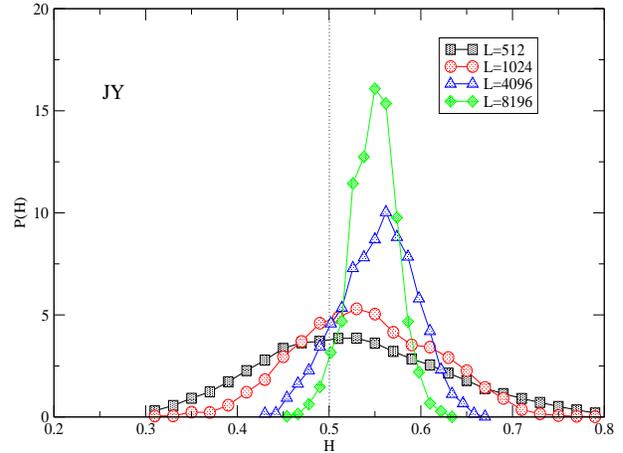}} \caption{For
the Japanese Yen futures the situation is very similar to the British Pound futures,
Fig.~\ref{fig::scaling_BP}, where overall persistent behaviour is evident. Different changes in the
dynamics are observed.} \label{fig::scaling_YJ}
\end{figure}

\begin{figure}
\vspace{1cm}
\centerline{\epsfig{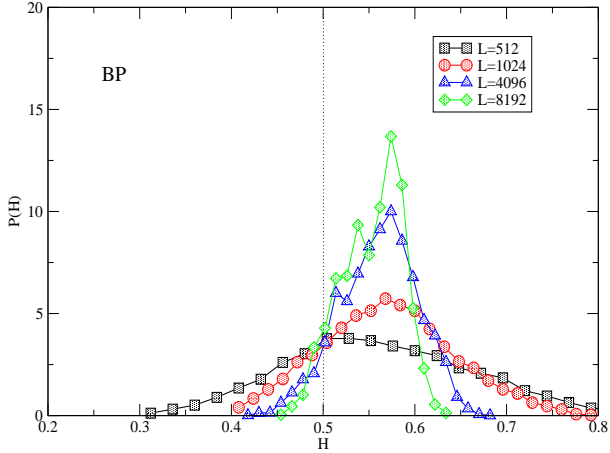}}
\caption{The British Pound futures display generally persistent
behaviour interrupted by many shoulders: different phases have
characterized the two years period considered.}
\label{fig::scaling_BP}
\end{figure}

\begin{figure}
\vspace{1cm} \centerline{\epsfig{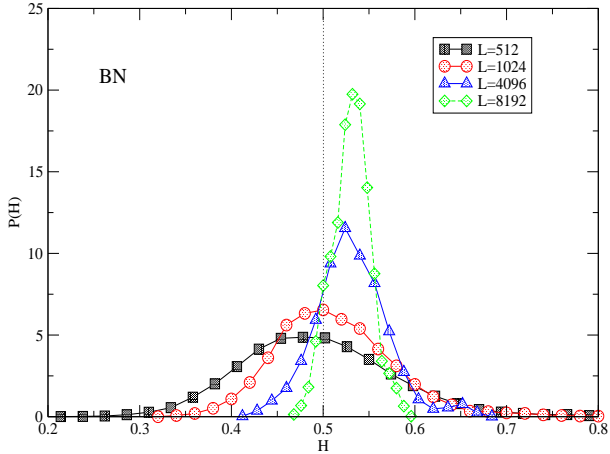}} \caption{The
pdfs for the Eurex Bunds futures show a quite singular behaviour. The index seems to shift from a
slightly persistent to a slightly antipersistent behaviour as we move toward smaller scales.}
\label{fig::scaling_BN}
\end{figure}

\begin{figure}
\vspace{1cm} \centerline{\epsfig{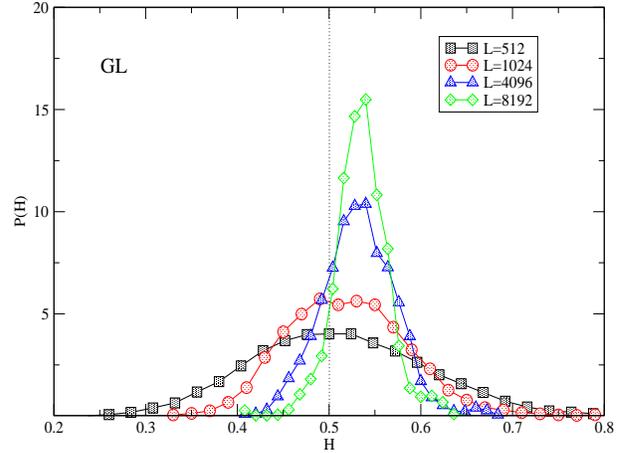}} \caption{The
pdfs for the Long Gilts futures point out an average slightly persistent behaviour until a time
frame of approximately one trading day ($L=512$). Shoulders are present as well.}
\label{fig::scaling_GL}
\end{figure}

\begin{figure}
\vspace{1cm} \centerline{\epsfig{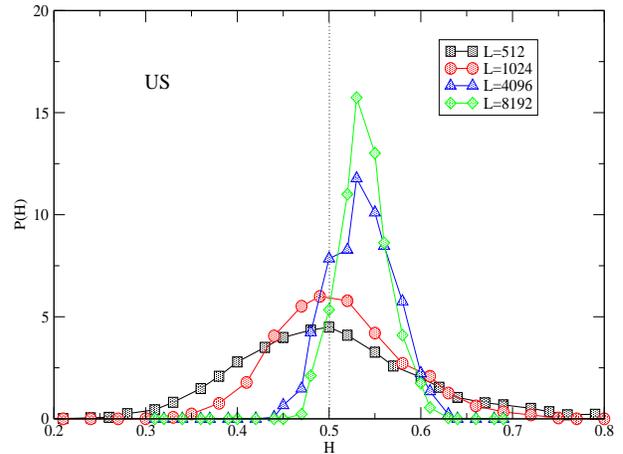}} \caption{Pdfs
for the Treasury Bonds futures. A smooth shift toward antipersistency at short time scales can be
observed.} \label{fig::scaling_US}
\end{figure}

\begin{figure}
\vspace{1cm} \centerline{\epsfig{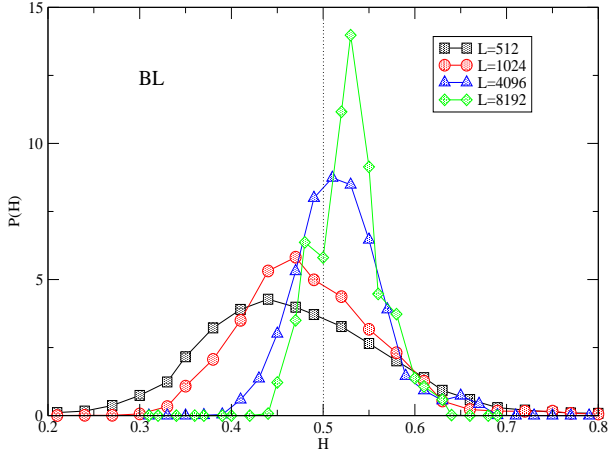}} \caption{Pdfs
for the BOBL futures. As for the Bunds, Fig.~\ref{fig::scaling_BN}, a crossover from persistency to
antipersistency is evident.} \label{fig::scaling_BL}
\end{figure}

In order to gain some insight into the general behaviour of the Hurst exponent at different scales
we now calculate the pdfs of $H_{L}(t)$ for various $L$. The results of the analysis are reported
from Fig.~\ref{fig::scaling_SP} to \ref{fig::scaling_BL}.

From these plots we can infer some interesting features which
characterized the multiscale dynamics of the indices under
investigation during the two years period analyzed.

First of all, the various distributions, irrespective of the particular index, are all centered not
too far from $0.5$, as we might have expected. In fact, large deviations from this average value
would have been related to significant inefficiencies which could have been subjected to market
arbitrage. These inefficiencies do not last for long in well developed markets such as the ones
studied in the present work.

As we examine the distributions in more detail we can notice that they get sharper as we increase
the length of the time scale, that is, the fluctuations around the average value decrease. This is
a natural consequence of the DFA-$2$ algorithm which produces a smaller dispersion on the value of
$H_{L}(t)$ as we increase the length of the time series to be evaluated.

More interestingly, a relatively smooth shift in the peak of the pdfs is evident as we move from
longer to shorter scales - this is particularly noticeable for the BP, BN, QM, BL, US, GL, JY, DA
and NK. Remarkably, in the BN, BL and US (fixed income) we observe a clear crossover from an
average persistent to an average antipersistent dynamics at a time scale of approximately 1 day. On
the other hand the HI displays a more persistent behaviour, on average, at shorter time scales.

For most of the indices there is also evidence of shoulders. These
anomalies indicate  that the system has moved through different
phases in the period under consideration and that in each phase
the Hurst exponent was significantly different from its average
value. This could be due to different exogenous reasons such as
changes in financial regulations, market ``mood" or just
variations in the trading mechanism.

\begin{figure}
\centerline{\epsfig{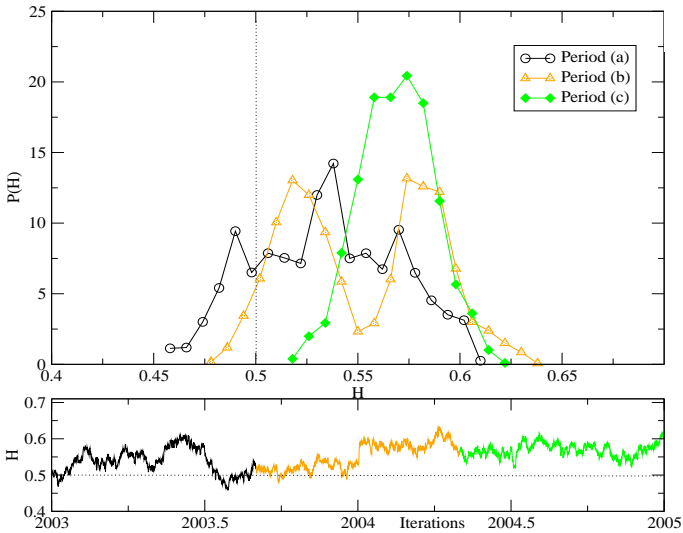}}
\caption{(Top) Pdfs of three subperiods, (Bottom), for the BP.
Note the different distributions of the Hurst exponent: these are
clear indications of non-stationarity in  the time series. In this
plot we used $L=8192$.} \label{fig:BP_split}
\end{figure}

In order to have a feeling of how the dynamics of an index can
differ over the two years period under consideration and to gain a
better insight into the origin of the shoulders we split the time
series of local Hurst exponent for the BP  into three identical
subperiods. Each of the subperiods corresponds to eight months of
trading. It is clear from the relative pdfs,
Fig.~\ref{fig:BP_split} (Top), that the dynamics of the system is
rather different in each of the subperiods and it becomes clear
how the shoulders have origins in these phases. Of course, this is
just a pedagogical example and, in principle, we could perform a
more accurate analysis by monitoring the changes in the pdf of
$H_{L}(t)$ in real time, as we did for the local Hurst exponent.
It would be interesting, for example, to monitor the behaviour of
the distribution of Hurst exponent at different scales just before
a market crash. However this issue goes beyond the aim of the
present paper and it will be addressed in a future work.

\begin{figure}
\vspace{1cm} \centerline{\epsfig{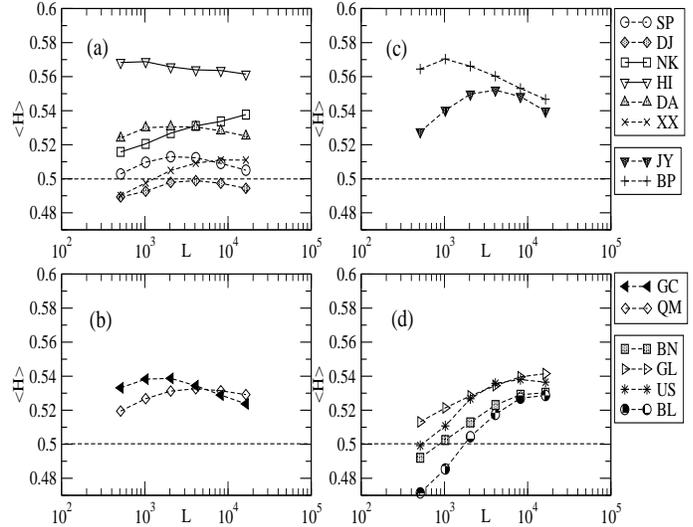}} \caption{Average
value of the local Hurst exponent, $\langle H \rangle_L$, for index futures (a), commodity futures
(b), exchange rate futures (c) and fixed income futures (d). The error bars on these points, not
plotted for clarity, are approximately equal to the standard deviations in Tab.~\ref{tab::bias}.
The horizontal dotted line is set at 0.5 for visual reference.} \label{fig::H_vs_L}
\end{figure}

The relationship between the market dynamics and the scale of observation appears to become more
evident when we plot the average value of the local Hurst exponent, $\langle H \rangle_L$, against
the scale $L$, Fig.~\ref{fig::H_vs_L}. From this graph we can notice how time series belonging to
the same sector tend to have a qualitatively similar scale dependency. The indices futures, for
example, Fig.~\ref{fig::H_vs_L} (a), do not display a strong correlation between $\langle H
\rangle_L$ and $L$ with the exception of the Hang Seng (HI) whose persistency increases sharply at
smaller scales. On the other hand a scale dependency is quite evident for the fixed income
products, Fig.~\ref{fig::H_vs_L} (d), where, interestingly, some time series (BN, US and BL) move
from an antipersistent-like to an persistent-like behaviour as the scale increases, as already
pointed out by our qualitative observation of the pdfs.

As a result of the analysis carried out in this section we can
claim that the actual behaviour of the stock market, described in
terms of Hurst exponent, apart from being influenced by the
particular period of time under consideration and by the maturity
of the
market~\cite{Costa03,DiMatteo03,Cajueiro04,DiMatteo05,Liu07}, is
also related to the particular scale of observation: this is an
extremely relevant issue for practical applications. In fact, if
we consider long time scales (large $L$), in reality, we are
estimating the {\em average} Hurst exponent over that period.

Moreover, these empirical findings confirm the multiscale and non-stationary nature of the stock
market, in contrast with the assumptions inherent in the EMH.

\subsection{Hurst exponent and end-of-day gaps}

Before concluding, we want also to briefly address the question of the relevance of end-of-day (EOD)
gaps in the analysis carried out so far. It is well known that the price can undergo large changes
while a market is not in session. This phenomenon is mainly related to the flow of information from
active markets in different time zones which, somehow, is ``digested"  by other markets during
their closure and then reflected in the opening price of the following day.

\begin{figure}
\vspace{1cm} \centerline{\epsfig{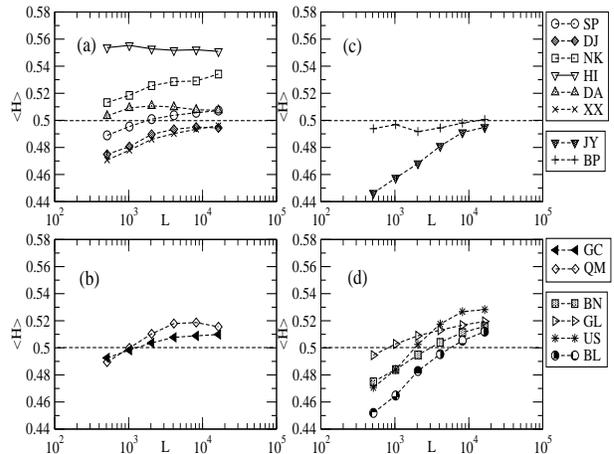}}
\caption{Same as Fig.~\ref{fig::H_vs_L} but, this time, in the analysis we do not consider price
returns that cross two different days.} \label{fig::H_vs_L_noEOD}
\end{figure}

So far we have considered these changes as a natural part of the dynamics of the market itself. In
this last section, instead, we want to consider the relative importance of these events in our
analysis by treating them as ``spurious" effects.
In fact, as we already discussed, large fluctuations tend to increase the value of the Hurst
exponent despite the genuine presence of persistency in the time series.
In order to account for this fact we have removed from the original time series the EOD returns and
then we have repeated the analysis performed in the previous section. A summary of the results for the
average value of $H_L(t)$ is shown in Fig.~\ref{fig::H_vs_L_noEOD}.  If we compare these plots with
Fig.~\ref{fig::H_vs_L} we can notice how most of the curves, while maintaining the same qualitative
shape, are shifted toward smaller values of $\langle H \rangle$, as expected.

However, it is important to point out how this shift can change the conclusions of the analysis in
terms of persistency/antipersistency. For example, the exchange rates futures appear to move from
an average persistent behaviour at all scales, Fig.~\ref{fig::H_vs_L} (c), to an average
uncorrelated behaviour for the BP and anticorrelated for the JY, Fig.~\ref{fig::H_vs_L_noEOD} (c).

This last analysis, therefore, confirms the relevance of the large fluctuations in the calculation
of Hurst exponents. Moreover, it points out the issue of the pre-conditioning of high frequency
data: this can lead to different conclusions regarding their behaviour across time series.

\section{Discussion and conclusion}

In the present work we have used the concept of local Hurst exponent in order to investigate the short
scale dynamical properties of the correlations in different future contracts (indices, commodities,
exchange rate and fixed income) from the beginning of 2003 to the end of 2004.

Analysis on the behaviour of $H_{L}(t)$ at different scales, and
in particular its distribution, points out a scale dependent and
non-stationary evolution of this scaling exponent, independent of
the specific kind of contract.

The Eurex Bunds, BOBL and U.S Treasury Bonds, for example, display
an average persistent behaviour over time scales of approximately
three weeks but they then become antipersistent, on average, for
time scales of the order of one day. Moreover, we observe changes
in the shape of the pdfs of $H_{L}(t)$ with time. This fact points
to the existence of different market phases in the two years
period from 1/1/2003 to 31/12/2004 and, therefore, evidence for
non-stationarity.
These empirical facts are in contrast with the EMH hypothesis,
according to which $H_{L}(t)$ should be constant and equal to 0.5
for each time scale.

It is worth to stress that the dynamical behaviour of the Hurst exponent is not related only to the
liquidity of the market but also to the variety of time horizons involved in the trade of a
particular asset. As a consequence, markets which involve many exogenous agents, such as the S\&P500,
tend to be more ``efficient".

In conclusion, we have shown that the concept of Hurst exponent for non-stationary time series has
a practical validity only in the period and the scale of observation. By estimating $H$ with a
large sample, due to the coarse-grain procedure of the DFA-$p$ algorithm, we lose the local
information and we obtain an ``average" value over that period. This can or cannot be a problem for
technical trading: it depends on the horizon we are interested in.
Moreover, market models  should be bounded to reproduce the time-scale variability of the Hurst
exponent, as already pointed out in Ref.~\cite{Costa03}.


In addition, we have shown that self-similarity of the large
fluctuations responsible for the ``fat" tail property of financial
time series can produce a substantial contribution to the value of
$H$ which is not related to temporal correlations in the price
variation. This effect cannot be neglected for high frequency
financial applications or, in general, for ``fat" tailed data
sets. In particular, we have pointed out the contribution of the
EOD gaps present in high frequency financial data. These results
are in agreement with the conclusions of Alfi et al.~\cite{Alfi07}
who investigated the robustness of the R/S
algorithm~\cite{Feder88} against tick-by-tick data from the New
York Stock Exchange. Note, however, that the DFA-$p$ algorithm
results to be more robust than the R/S method with respect to both
the large fluctuations and the window size. In particular, for
small samples and Gaussian increments, we do not observe the
systematic bias of $H$ that was reported in~\cite{Alfi07} - at
least for $0.3 \lesssim H \lesssim 0.8$ (see
Sec.~\ref{sec::gaussian_inc} for details).

Given the present results we feel it will be necessary to develop
an alternative methodology - one that is not based on scaling
exponents - if we are to reliably exploit the high order
correlations found in non-stationary and fat tailed data sets.
Some alternatives could be found in random matrix
theory~\cite{Drozdz01,Plerou02,Potters05}, hyperbolic
networks~\cite{Tumminiello05}, or information theory tools such as
the transfer entropy~\cite{Marschinski02}.

Finally, it is worth noting that the time/scale dependency of the
scaling exponent $H$ investigated in the present work can also be
extended to the multifractal framework~\cite{Kantelhardt02}. In
this case, the time series is assumed to be characterized not by
one but by an entire spectrum of scaling exponents. Our future
work will be devoted to the study of the temporal properties of
these multifractal spectra in different financial time
series~\cite{Bouchaud00,Lux01,Calvet02,Lux04,Borland05,DiMatteo07}
along with their financial implications.

\section*{Acknowledgement}
The authors would like to thank Richard Grinham for many useful
discussions and a careful reading of the manuscript. TDM and TA
acknowledge the partial support by ARC Discovery Projects:
DP03440044 (2003) and DP0558183 (2005), COST P10 ``Physics of
Risk" project and M.I.U.R.-F.I.S.R. Project ``Ultra-high frequency
dynamics of financial markets".

\end{document}